\documentclass[twocolumn,showpacs,prb,graphics,graphicx]{revtex4-1}
\usepackage{bm}
\usepackage[pdftex]{graphicx}
\usepackage{dcolumn}

\begin{document}

\title{Elastic contribution to interaction of   vortices in uniaxial  superconductors}
\author{V. G. Kogan}
\affiliation{Ames Laboratory - DOE and Department of Physics, Iowa State University,
Ames, IA 50011}

\begin{abstract}
The stress caused by vortices in  tetragonal superconductors    contributes to the intervortex interaction which depends on vortex orientation within the crystal, on elastic moduli, and is attractive within certain angular regions even in fields along the $c$ crystal axis.     For sufficiently strong stress dependence of the critical temperature,  this contribution may result in distortions of the hexagonal vortex lattice for $\bm H||\bm c$.   In small fields it leads to  
formation of a square vortex lattice with a fixed $H$ independent spacing. This should be seen in the   magnetization $M(H)$ as a discontinuous jump of magnetization at the transition from the Meissner to mixed states.
\end{abstract}

\date{\today}

 \pacs{74.20.De,74.25.Ha,74.25.Wx,74.62.Fj}
\maketitle

The strains in type-II superconductors that arise due to presence of vortices and defects were considered time ago as a pinning mechanism, see, e.g., Ref.\,\onlinecite{Campbell-Evetts}  and references therein. Recently, vortex induced  strains, their weakness notwithstanding, were shown to cause substantial changes in macroscopic magnetization  in materials with strong pressure dependence of the critical temperature $T_c$,\cite{RapCom}  the result of a long-range  elastic perturbations as opposed to a short-range London intervortex interactions.
Here,  uniaxial materials are  considered in  fields along  principal crystal directions. 
To calculate the elastic energy, vortices are treated   as one-dimensional strain sources in an infinite crystal, so that we deal with  a  {\it planar} elasticity  problem. \cite{LL}

The value of  $dT_c/dp$ ($p$ is the stress  or, in particular, the pressure) is an indicator of strength of magneto-elastic effects in the mixed state. It turned out recently that  this derivative in pnictides, and in Ca(Fe$_{1-x}$Co$_x$)$_2$As$_2$ in particular,\cite{dTc/dp} by one or two orders of magnitude exceeds   values for conventional superconductors making Fe-based materials  favorable for observation of   magneto-elastic effects.

It is shown below that elastic  intervortex interactions in tetragonal materials are strongly anisotropic even  for vortices  directed along the $c$ crystal axis. For a particular set of elastic moduli, two vortices situated at [100] or [010] elastically attract each other, whereas being at [110] they are repelled. 
 In large fields, the extra interaction removes orientational degeneracy of the standard $60^\circ$ triangular vortex lattice and should cause distortions of this structure  in qualitative agreement with  data on KFe$_2$As$_2$.\cite{KFe2As2}   Possible relevance of the elasticity in hexagonal crystals for low fields vortex arrangement in MgB$_2$ is discussed.\cite{Mosch3}\\

%%%%%%
%\section{Anisotropic crystal}
%%%%%%%

 1. {\bf Tetragonal crystal}.  The general form of elastic energy density is:\cite{LL}
\begin{eqnarray}
F= \lambda_{iklm}u_{ik} u_{lm} /2 \,.
 \label{eq1}
\end{eqnarray}
where $u_{ik}$ are strains and $\lambda_{iklm}$ are elastic moduli. For brevity we denote the non-zero components of the elastic tensor in the crystal frame as:\cite{remark1}
\begin{eqnarray}
 \lambda_{aaaa}&=& \lambda_{bbbb}=\lambda_1  \,,\quad  \lambda_{aabb}=\lambda_2, \quad \lambda_{abab}=\lambda_3\,,\nonumber\\
\lambda_{cccc} &=&\lambda_4\,,\quad  \lambda_{aacc}=\lambda_{bbcc}= \lambda_5,\nonumber\\   
 \lambda_{acac} &=&   \lambda_{bcbc}  = \lambda_6\,.
  \label{moduli}
\end{eqnarray}

 Consider a system of straight parallel vortices along $z$ tilted with respect to the frame $(a,b,c)$ of an infinite  tetragonal crystal.  Introduce the vortex frame  $(x,y,z)$  so that   the angle between the vortex axis $z$ and the $c$ axis  is $\theta$. Consider the tilt as rotation round the $a$ axis:
\begin{eqnarray} x=a,\quad
 y =b\cos\theta-c \sin\theta \, ,\quad z  =b\sin\theta+c \cos\theta\,. \qquad  
  \label{C27}
\end{eqnarray}
Although this a particular tilt, the resulting physics only weakly depends on this choice. 

The elastic perturbation  by vortices is caused by a number of reasons among which  effects of normal cores   and of  supercurrents around  were  discussed, see, e.g., Refs.\,\onlinecite{KBMD} and \onlinecite{Cano} and references therein. In this communication, only {\it planar} deformations are considered, i.e., such that  
the displacement  $u_z$ is zero and all $u_{iz}=0$.   

Components of the stress tensor   $\sigma_{ik}=\lambda_{iklm}  u_{lm}=\lambda_{ik\alpha\beta}  u_{\alpha\beta}$ (hereafter Greek indices take only $x,y$ values)  are:
 \begin{eqnarray}
\sigma_{xx}&=&\lambda_{xxxx}  u_{xx}+ \lambda_{xxyy}  u_{yy} \,,\label{s-xx}\\
\sigma_{yy}&=&\lambda_{yyxx}  u_{xx} + \lambda_{yyyy}  u_{yy}\,,\label{s-yy} \\ 
\sigma_{xy}&=&2 \lambda_{xyxy}  u_{xy} \,,\label{s-xy}
\end{eqnarray}
and
 \begin{eqnarray}
\sigma_{zz}&=&\lambda_{zzxx}  u_{xx}+ \lambda_{zzyy}  u_{yy} \,.
 \label{eq-6}
\end{eqnarray}
Further, 
 \begin{eqnarray}
\sigma_{xz}=2\lambda_{xzxy}  u_{xy}  \,,\qquad
\sigma_{yz}=\lambda_{yzxx}  u_{xx} + \lambda_{yzyy}  u_{yy}\,.\qquad
\label{s-yz}   
\end{eqnarray}
    The  needed elastic tensor  components in the vortex frame are given in Appendix A.

Given only two independent displacements $u_x, u_y$ in the {\it planar} problem of  our  interest, the components of the stress tensor cannot be independent.\cite{LL} 
Indeed,  express $u_{xx}$ and $u_{yy}$ from Eqs.\, (\ref{s-xx}) and  (\ref{s-yy}) 
  \begin{eqnarray}
 u_{xx}\,d&=& \lambda_{yyyy}  \sigma_{xx}-\lambda_{xxyy}  \sigma_{yy}  \,,\label{uxx}\\
 u_{yy}\,d&=& \lambda_{1}  \sigma_{yy}-\lambda_{xxyy}  \sigma_{xx} \,, \label{uyy}\\
  d&=&\lambda_1 \lambda_{yyyy} - \lambda_{xxyy}^2 \,,   \label{C2}
 \end{eqnarray}
and substitute the result in Eq.\,(\ref{eq-6}):
 \begin{eqnarray}
  &&\sigma_{zz} d = d_1  \sigma_{xx}+d_2  \sigma_{yy}  \,,\label{eq8}\\
&& d_1= \lambda_{zzxx} \lambda_{yyyy}- \lambda_{zzyy} \lambda_{xxyy}   \,, \label{d1}\\ 
&&d_2= \lambda_{zzyy} \lambda_{1}- \lambda_{zzxx} \lambda_{xxyy}   \,. \label{d2}
\end{eqnarray}

The equilibrium conditions 
 $ \partial \sigma_{\alpha\beta}/\partial x_\beta  =0$ or 
\begin{eqnarray}
\frac{\partial \sigma_{xx}}{\partial x}+  \frac{\partial \sigma_{xy}}{\partial y}=0\,,\qquad  \frac{\partial \sigma_{yx}}{\partial x}+  \frac{\partial \sigma_{yy}}{\partial y}=0\,, 
 \label{equilibrium}
\end{eqnarray}
are satisfied if one sets
\begin{eqnarray}
 \sigma_{xx}=\frac{\partial^2\chi}{\partial y^2}  ,\quad \sigma_{yy}=\frac{\partial^2\chi}{\partial x^2}  ,\quad \sigma_{xy}=-\frac{\partial^2\chi}{\partial x\partial y}  
   \label{C5}
\end{eqnarray}
with an arbitrary function $\chi(x,y)$.\cite{LL} Relation (\ref{eq8}) then provides an equation for $\chi$:
 \begin{eqnarray}
  d_1  \frac{\partial^2\chi}{\partial y^2}+d_2 \frac{\partial^2\chi}{\partial x^2} =  d\,\sigma_{zz}  \,.\label{eq11}
\end{eqnarray}
 Since $\sigma_{ii}=-3p$ with the pressure $p$,   we have
 \begin{eqnarray}
  \sigma_{zz}  = -3p- \nabla^2\chi\,.
  \label{eq12}
\end{eqnarray}
Hence,  we obtain:
 \begin{eqnarray}
&&D_2 \frac{\partial^2\chi}{\partial x^2}+D_1  \frac{\partial^2\chi}{\partial y^2}  = -3 pd  \,.\label{eq13}\\
&&  D_1=d+d_1\,,\qquad D_2=d+d_2.\qquad\label{D1D2}
\end{eqnarray}
 The rescaling  
  \begin{eqnarray}
  x_n=x \sqrt{ d/D_2 }\,,\qquad  y_n=y\sqrt{ d/D_1} \label{rescale}
  \end{eqnarray} 
  transforms this to Poisson  equation which can be solved for $\chi(x,y)$. Hence, both the  stresses $\sigma_{\alpha\beta}$ and strains  $u_{\alpha\beta}$ can be found.  
  
  At  first sight, the problem of elastic perturbation caused by parallel vortices  can be considered as planar for any orientation of vortices relative to the crystal. This is, however, not the case. The point is that in general the equilibrium conditions (\ref{equilibrium}) should be complemented by  $ \partial \sigma_{z\beta}/\partial x_\beta  =0$ with $\sigma_{z\beta}$ given in Eq.\.(\ref{s-yz}). However, since the planar solutions  $u_{\alpha\beta}$ are already fixed to satisfy (\ref{equilibrium}), there is no room for any extra conditions.  The only situation free of this contradiction is when both $\sigma_{zx}$ and $\sigma_{zy}$ are zeros. The direct examination of elastic constants involved in $\sigma_{z\beta}$ shows that they are $\propto \sin\theta\cos\theta$, i.e., $\sigma_{z\beta}=0$   only for   $\theta=0$ and $\theta=\pi/2$. Since the problem is planar indeed for these orientations, they are  considered in what follows.\\

%%%%%%%
%\subsection{Single vortex}
%%%%%%%
   
  2. {\bf Single vortex},   $\bm {\theta=0}$.  
    For $\theta=0$  we have $\lambda_{yyyy}=\lambda_1$, $\lambda_{xxyy}=\lambda_2$, $\lambda_{xyxy}=\lambda_3$, $\lambda_{xxzz}=\lambda_{yyzz}=\lambda_5$. Further, $d=\lambda_1^2-\lambda_2^2$, $d_1=d_2=\lambda_5(\lambda_1-\lambda_2)$, and   $D_1=D_2=D=(\lambda_1-\lambda_2)(\lambda_1+\lambda_2+\lambda_5)$. 

  For a single vortex in infinite sample the pressure is zero, whereas the stress due to the vortex can be described as due to a $\delta$-function singular source. We have instead of Eq.\,(\ref{eq13}):
 \begin{eqnarray}
 \nabla^2\chi  =  2\pi S(0)\,\delta(\bm r-\bm a)  \,,
  \label{Poisson1}
\end{eqnarray}
where  $d/D\sim 1$ is incorporated in $S$. Note that 
no vortex   model is used  here explicitly; the vortex is described  as a $\delta$-function stress source of  a ``strength" $S$ which can be estimated by a  better-than-London theory. This approach is justified since the elastic perturbation is long-range ($\propto 1/r^2$); hence both the effect of the core of a  size $\xi$ and of out-of-core region\cite{Cano,Cano1}  on the order of the London penetration depth can be  included in the point source. 

The solution is $\chi=S(0)\ln |\bm r-\bm a|$ or in the Fourier space:
\begin{eqnarray}
\chi (\bm k)= -\frac{2\pi S(0)}{k^2}\,e^{i{\bm k}{\bm a} }. 
 \label{30}
\end{eqnarray}
Hence we have:
\begin{eqnarray}
\sigma_{xx} (\bm k)=  \frac{2\pi  S(0)k_y^2}{ k^2}\,e^{i\bm {ka}},\,\,\, \sigma_{yy} (\bm k)=   \frac{2\pi  S(0)k_y^2}{ k^2}\,e^{i\bm {ka}},\nonumber\\
\sigma_{xy} (\bm k)= -  \frac{2\pi  S(0)k_xk_y^2}{ k^2}\,e^{i\bm {ka}}, \qquad
 \label{31}
\end{eqnarray}
and with the help of Eqs.\,(\ref{uxx}), (\ref{uyy}), and (\ref{s-xy})
\begin{eqnarray}
u_{xx} &=&    \frac{2\pi  S(0) }{ d}\,
\frac{\lambda_1  k_y^2-\lambda_2  k_x^2}{k^2} e^{i\bm {ka}},\qquad\label{uxx(k)}\\
 u_{yy} &=& \frac{2\pi  S(0) }{ d}\,
\frac{\lambda_1  k_x^2-\lambda_2  k_y^2}{k^2} e^{i\bm {ka}},\qquad \label{uyy(k)}\\
u_{xy} &=&  \frac{ \pi  S(0) }{\lambda_3}\,
\frac{   k_x   k_y }{k^2} e^{i\bm {ka}}\,. \label{uxy(k)}
 \label{31}
\end{eqnarray}

%%%%%%%%%%
%%%%%%%%%%%
 3. {\bf Elastic    interaction of vortices}, $ \bm {\theta=0}$.  Let a vortex be at the origin and  another one at $\bm a$. The elastic energy ${\cal E}=\int d{\bm r} \sigma_{\alpha\beta}u_{\alpha\beta}/2$   contains the self-energies of each one of them and   the interaction energy:
\begin{eqnarray}
  {\cal E}_{int}
%=\frac{1}{2} \int d{\bm r}  \left[\sigma(0)_{\alpha\beta}\, u_{\alpha\beta}(\bm a) +\sigma(\bm a)_{\alpha\beta}\, u_{\alpha\beta}(0)\right]  \nonumber\\
= \int d{\bm r}   \sigma(0)_{\alpha\beta}\, u_{\alpha\beta}(\bm a)= \int \frac{d{\bm k}}{4\pi^2}   \sigma_{\alpha\beta}(0,\bm k)  \, u_{\alpha\beta}(\bm a,-\bm  k) .\nonumber\\
 \label{int}
\end{eqnarray}
After straightforward algebra one obtains for $\theta=0$:
\begin{eqnarray}
\frac{{\cal E}_{int}}{ S^2(0)} = \frac{d}{D^2}\left[ \lambda_1(I_1+   I_2) +  \frac{ d-2 \lambda_2 \lambda_3 }{\lambda_3 }I_3\right] \qquad
 \label{interaction}
\end{eqnarray}
where 
\begin{eqnarray}
 I_1  &=&\int  d{\bm k}   \frac{k_x^4}{k^4}\,e^{-i\bm {ka} },\quad  I_2=\int  d{\bm k}   \frac{k_y^4}{k^4}\,e^{-i\bm {ka} },\nonumber\\
   I_3&=&\int  d{\bm k}   \frac{k_x^2k_y^2}{k^4}\,e^{-i\bm {ka} },\qquad
 \label{int3}
\end{eqnarray}
%where the subscript $n$ for rescaled positions ${\bm a}_n$ is omitted for brevity.  
One easily verifies that $I_1+I_2=-2I_3$ if $\bm a\ne 0$. Thus,  for $\theta=0$ the interaction energy is proportional to $I_3(\bm a)$. 
 
Consider the second vortex in the first quadrant $a_x>0, a_y>0$. 
To evaluate, e.g., $I_2$,   integrate first over $k_x$ utilizing the pole $k_x=-i|k_y|$ in the lower half of the complex plane $k_x$:
\begin{eqnarray}
\int_{-\infty}^\infty  \frac{dk_x e^{-ik_xa_x}}{(k_x^2+k_y^2)^2}= \frac{\pi}{2k_y^4}e^{-|k_y|a_x}(a_xk_y^2+|k_y|).
 \label{int2}
\end{eqnarray}
Integration over $k_y$ gives:
 \begin{eqnarray}
I_2 =   \pi\frac{3a_x^4-6a_x^2a_y^2-a_y^4}{a^6}.
 \label{int_xx}
\end{eqnarray}
Similarly one obtains 
\begin{eqnarray}
I_1 =   \pi\frac{3a_y^4-6a_x^2a_y^2-a_x^4}{a^6},\,\,\,\, 
I_3 =  - \pi\frac{a_x^4-6a_x^2a_y^2+a_y^4}{a^6}. \qquad
 \label{int3}
\end{eqnarray}
In polar coordinates $a_x=a\cos\varphi, a_y=a\sin\varphi$, one has:
\begin{eqnarray}
I_3 &=&  - \frac{\pi}{a^2}\,  \cos 4\varphi \,,\quad I_2 =  \frac{\pi}{a^2}\, (\cos 4\varphi +2\cos 2\varphi),\nonumber\\
I_1 &=&   \frac{\pi}{a^2}\, (\cos 4\varphi -2\cos 2\varphi).  
  \label{I(phi)}
\end{eqnarray}
 \begin{figure}[h]
\includegraphics[width=8cm]{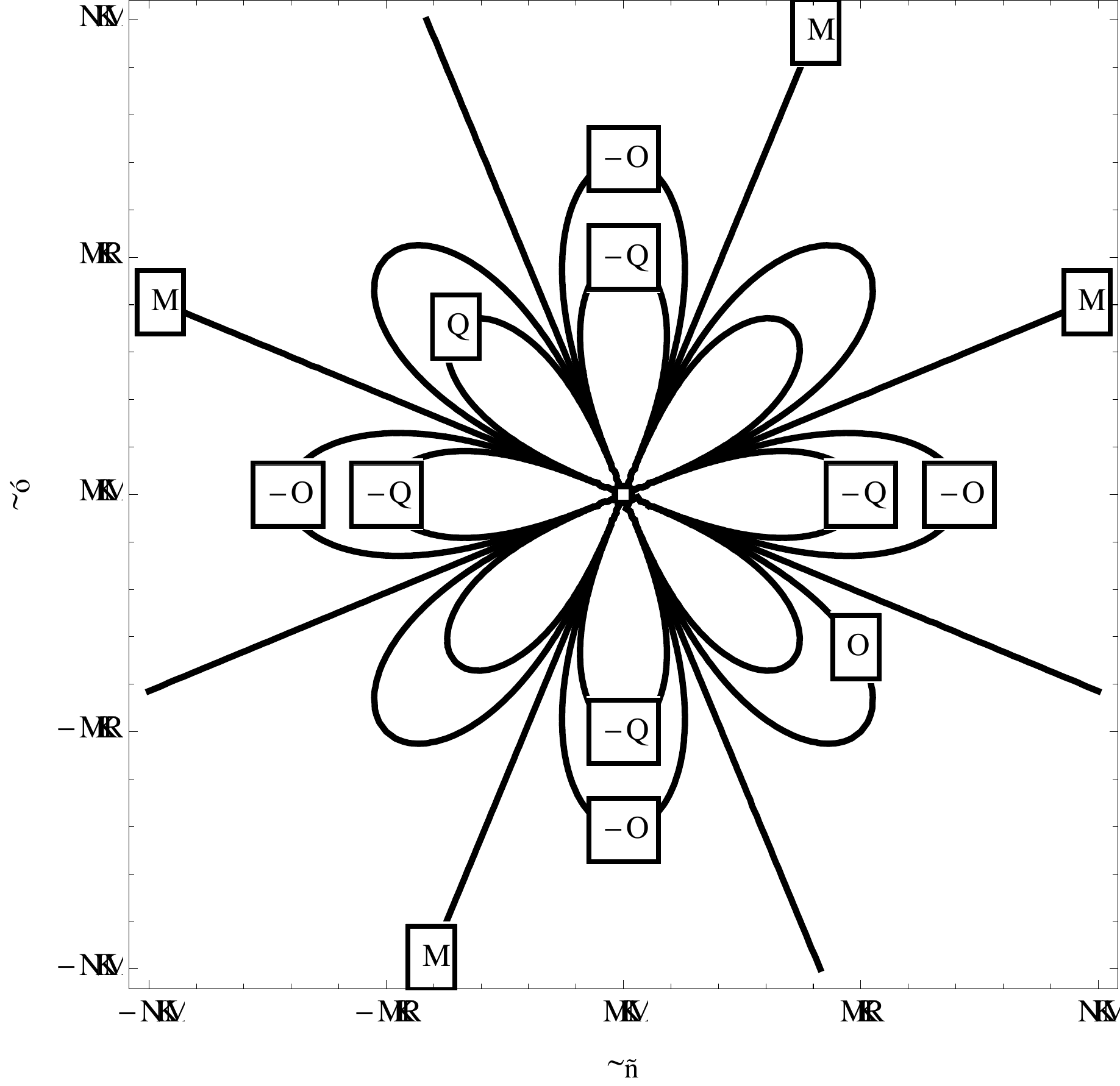}
 \caption{  Contours of constant elastic interaction energy or of  $I_3(a_x,a_y) $      corresponding to a vortex at the origin and another one at $\bm a =(a_x,a_y)$. Both vortices are parallel to the $c$ axis.  $\lambda_1- \lambda_2 -2\lambda_3 $ is assumed positive. 
 }
 \label{fig1}
 \end{figure}

Hence, the interaction energy takes the form:
\begin{eqnarray}
\frac{{\cal E}_{int}}{ S^2(0)} =- \frac{\pi d^3}{D^4\lambda_3} \,  \frac{ (\lambda_1+\lambda_2)(\lambda_1- \lambda_2 -2\lambda_3) }{a^2 }\cos 4\varphi .\qquad
 \label{interaction1}
\end{eqnarray}
The energy ${\cal E}_{int}$ changes sign at $\varphi=\pi/8$ and $3\pi/8$ in the first quadrant. If $\lambda_1- \lambda_2 -2\lambda_3>0$ and the intervortex vector $\bm a$ is  in the  domain $-\pi/8<\varphi<\pi/8$ adjacent to [100], the elastic contribution to the  interaction is attractive, whereas it is repulsive for $\pi/8<\varphi<3\pi/8$ near [110]. 
Fig.\,\ref{fig1} shows contours of a constant $I_3(a_x,a_y)$. Note  that  ${\cal E}_{int} $ in addition to either attractive or repulsive
dependence on the intervortex distance $a$, depends on the azimuth $\varphi$, which for the example shown in the figure means that for a given $a$ the second vortex is pushed toward [100] or [010]. \\

%%%%%%%%%
3. {\bf Elastic    interaction}, $\bm{\theta=\pi/2}$. 
%%%%%%%%

In this case, examination of elastic moduli, Appedix A, gives $\lambda_{yyyy}=\lambda_4$, $\lambda_{xxyy}=\lambda_5$, $\lambda_{xyxy}=\lambda_6$, $\lambda_{xxzz}=\lambda_2$, $\lambda_{yyzz}=\lambda_5$. Further, $d=\lambda_1\lambda_4-\lambda_5^2$, $D_1=\lambda_4(\lambda_1+\lambda_2)-2\lambda_5^2$, and  $D_2=\lambda_1(\lambda_4+\lambda_5)-\lambda_5(\lambda_5+\lambda_2)$.
 
The same argument which led to  derivation of Eq.\,(\ref{Poisson1})  from Eq.\,(\ref{eq13}) leads again to a Poisson equation but in rescaled coordinates (\ref{rescale}) with, however, a new source $S(\pi/2)$. We then obtain the same expressions for $\sigma_{\alpha\beta}(\bm k)$ in properly rescaled $\bm k$'s and $\bm a$'s. The   strains are obtained with the help of   Eqs.\,(\ref{uxx}), (\ref{uyy}) and (\ref{s-xy}). 

The interaction energy now is a linear combination of $I_1, I_2$ and $I_3$ with the moduli dependent coefficients.  Without going to details, one can write this energy as:
\begin{eqnarray}
 {\cal E}_{int} = S^2(\pi/2)\,\frac {A \cos 4\varphi +B\cos 2\varphi  }{a^2} \,   
 \label{interaction2}
\end{eqnarray}
where $A,B$ depend on elastic moduli. The interaction energy may have a structure similar  to the case $\theta=0$ of Fig.\ref{fig1} but distorted due to the lost symmetry of the $90^\circ$ rotation.  
 \begin{figure}[h]
\includegraphics[width=8cm]{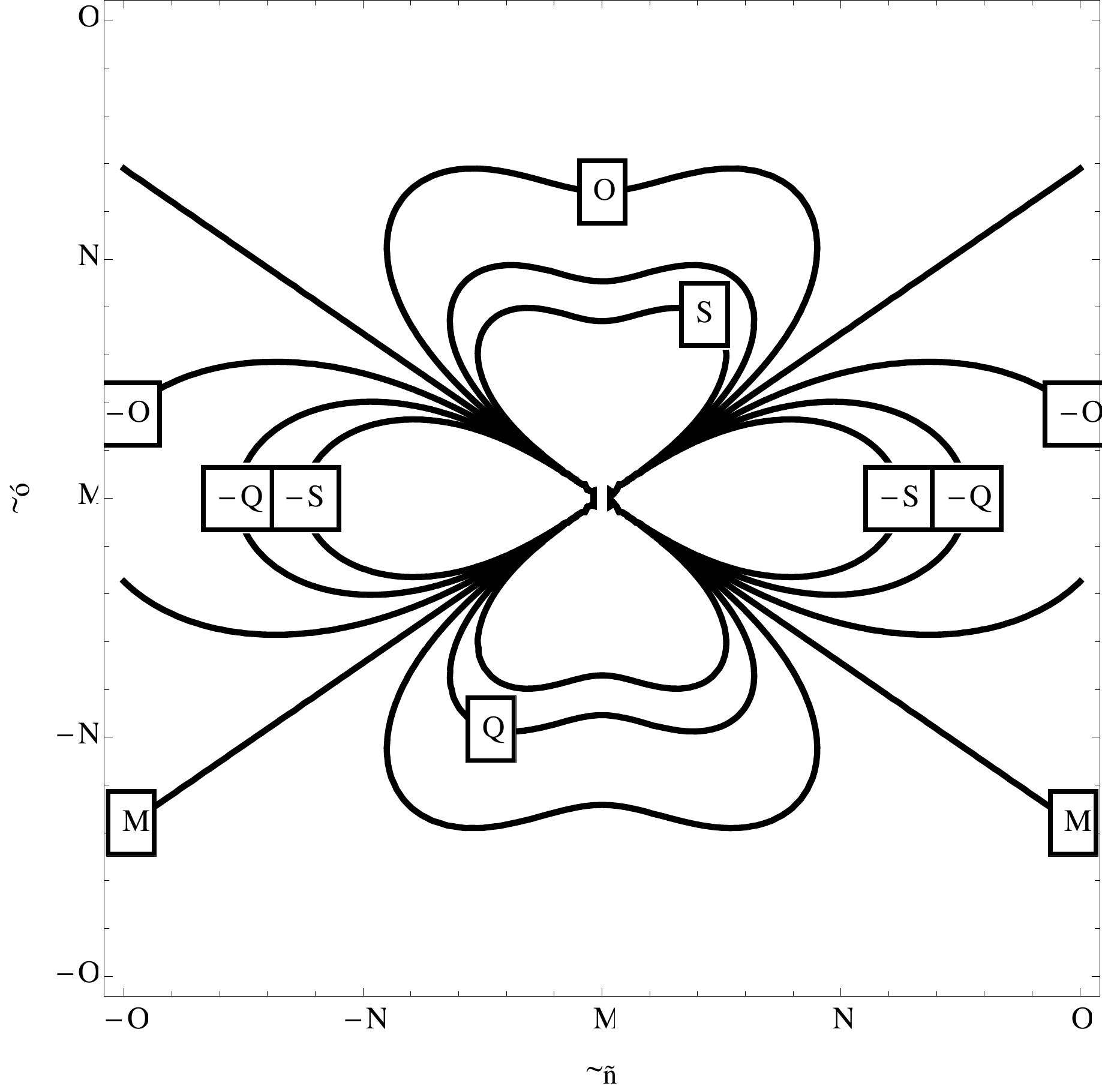}
 \caption{   Contours of constant elastic interaction energy  corresponding to a vortex at the origin and another one at a rescaled position $ (a_x, a_y)$. Both vortices are parallel to the $b$ crystal axis. The elastic moduli chosen are $\lambda_1=1, \lambda_2=0.5, \lambda_3=0.3, \lambda_4=0.2, \lambda_5=0.075, \lambda_6=0.2$.   }
 \label{fig2}
 \end{figure}
 In other words, attractions for $\bm a$ along [100] and [001] are no longer the same,   the attraction domain adjacent to [001] may shrink. For certain choices of moduli, this domain  disappears altogether, an example is shown in Fig.\,\ref{fig2} where [100] corresponds to attraction whereas [001] to repulsion. \\
  
%%%%%%%%
%%%%%%%%
  7. {\bf Discussion.} 
  The straightforward technique offered above can be applied to variety of situations for which   the elastic moduli are known. There is no point in going to all such possibilities in a general discussion of this paper. It is worth noting, however, that for vortices along the $c$ axis of tetragonal crystals, the generic form of this interaction is that of the geometric factor $I_3(a_x, a_y)$ with the sign and value of the pre-factor depending on elastic moduli, i.e., $ {\cal E}_{int}$ should always be of the form shown in Fig.\,\ref{fig1}. This implies in particular that  the elastic contribution removes the  degeneracy of orientations of the hexagonal vortex lattice for this field orientation. Moreover, 
 since at large distances elastic $1/a^2$ interaction   overcomes the exponentially weak London  repulsion, one expects  low density vortices parallel to $c$ to form a square lattice.   In small in-plane oriented  fields,  $\theta=\pi/2$, one expects to have vortex chains  along $b$ or $c$ depending on a  particular set  of elastic constants.  

  It is also worth noting that the method developed for tetragonal crystals can be applied 
  for the cubic symmetry by setting $\lambda_4=\lambda_1,  \lambda_5=\lambda_2$, and $\lambda_6=\lambda_3 $, see Eqs.(\ref{moduli}). It is also easily shown that the elastic energy for the hexagonal symmetry can be obtained from the tetragonal expressions by setting $\lambda_1=\lambda_2+2\lambda_6$, Appendix B. Then Eq.\,(\ref{interaction1}) shows  that vortices parallel to $c$ of a hexagonal crystal  do not interact elastically, the conclusion obtained in Ref.\,\onlinecite{KBMD}. 
  
As was shown above, the elastic field created by parallel vortices can be considered as {\it plane} only when vortices are oriented along  principal crystal directions. This is not so for a general vortex orientation, when the  strength $S$ of the stress  caused by vortex cores and by the currents around may depend on vortex orientation in a non-trivial manner; enough to mention that in this situation the currents do not flow exclusively in the plane perpendicular to the vortex direction. For arbitrary orientations,  
 one should use a more general approach  involving the Green's function of anisotropic elastic media.\cite{Lifshitz}  This, in fact, has been done for   hexagonal crystals.\cite{KBMD} This technique is formally  more involved  while, as far as physics is concerned, the major features of elastic interactions are already seen in the limiting cases of $\theta=0$ and $\pi/2$. 
  
  We now compare the elastic contribution with the standard London repulsion 
\begin{eqnarray}
 {\cal E}_L  =\frac{\phi_0^2}{8\pi^2\lambda_L^2} K_0\left(\frac{a}{\lambda_L}\right)\,,
 \label{London}
\end{eqnarray}
where   $\lambda_L$ is the London penetration depth. Since the $a$-dependent factors $I_{1,2,3}$ in the elastic interaction (\ref{interaction}) go as $1/a^2$, we estimate:
\begin{eqnarray}
 {\cal F}_{el} \sim   S^2/\tilde\lambda a^2   \,,
 \label{Fel}
\end{eqnarray}
where $\tilde\lambda\sim 10^{12}$erg/cm$^3$ is the order of magnitude of elastic constants. 
    At large  distances the power-law elastic interaction dominates the exponentially weak London repulsion. Hence,   $ {\cal E}_{int}$ positive at short distances, may turn negative along the directions where the elastic attraction exceeds the London repulsion, i.e., $ {\cal E}_{int}(a)$ goes through a minimum at some $a_m$, and approaches zero being negative as $a\to\infty$. In other words, in small fields vortices parallel to $c$ will occupy these minima, i.e.,  form a square lattice in the tetragonal or cubic cases with a fixed field-independent  spacing close to $a_m$. 

In Ref.\,\onlinecite{RapCom}, the source term in the equation for {\it strains}, $\nabla^2\chi_s=2\pi S_s\delta(\bm r-\bm a)$, due to the  vortex core was estimated as $S_s\sim \zeta \xi^2$, where $\xi$ is the core size and 
\begin{eqnarray}
\zeta \approx    \frac{H_c^2}{T_c} \,  \frac{dT_c}{dp}   
 \label{Fel}
\end{eqnarray}
is the relative volume change of normal and superconducting phases.
It was argued later that this estimate, based on the core solely responsible for the strain, underestimates the vortex induced strain by a factor as large as $(\ln\kappa)^2$ with $\kappa$ being the Ginzburg-Landau parameter.\cite{Cano,Cano1} Hence, the stress source  in our case can be estimated as
\begin{eqnarray}
S\sim \tilde\lambda\zeta\xi^2 (\ln\kappa)^2 .  \label{S-est}
\end{eqnarray}

The distance at which the elastic and London interactions are of the same order is estimated by setting ${\cal F}_{el}\approx {\cal F}_{L}$:
\begin{eqnarray}
 \frac{a^2}{ \lambda_L^2} K_0\left(\frac{a}{\lambda_L}\right)\approx\frac{S^2 }{\tilde\lambda \phi_0^2} \approx \frac{\tilde\lambda\zeta^2\xi^4 (\ln\kappa)^4}{\phi_0^2}\,.
 \label{eq for am}
\end{eqnarray}
Roughly one  estimates the right-hand side of this equation as $5\times 10^{-4}$ taking a moderate value $dT_c/dp\approx 1\,$K/GPa=$ 10^{-10}$K\,cm$^3$/erg  and  $H_c\approx 1\,$T. Solving Eq.\,(\ref{eq for am}) numerically, we get $a/\lambda_L\approx 10$. Taking this distance as a side $a_m$ of the low field square vortex lattice in tetragonal crystals we obtain the paramagnetic contribution to the magnetization  $\Delta M=\phi_0/4\pi a_m^2$. Therefore, at the transition from Meissner to the mixed state at the entry field $H_{ent}$, $M(H)$ should jump from $-H_{ent}/4\pi$ on the Meissner side to $-H_{ent}/4\pi+\Delta M$. For $\lambda_L\sim 10^{-5}\,$cm we estimate $\Delta M\sim 1\,$G. 

As mentioned, in some Fe-based compounds the derivative $dT_c/dp$ is   by an order of magnitude larger\cite{dTc/dp} which results in higher estimates for  $\Delta M $. This might be a reason for a very sharp and narrow peak in $M(H)$ at low fields observed in many compounds of this family.\cite{Proz,loops}  

As argued above, in hexagonal crystals the elastic interaction is absent for vortices directed along $\bm c$. However, a small misalignment of these directions may cause vortex chains to appear with a field independent intrachain  spacing. 
 It is of interest to compare our estimates with vortex chains in MgB$_2$ observed is small fields nominally parallel to $\bm c$.\cite{Mosch3} 
% Although MgB$_2$ is hexagonal, the above rough estimates should hold. 
 In this material $dT_c/dp\approx 1\,$K/GPa=$ 10^{-10}$K cm$^3$/erg.\cite{MgB2} Taking $H_c\approx 1\,$T, we obtain the same estimate as above: $a/\lambda_L\approx 10$. Ref.\,\onlinecite{Mosch3} reports a nearly field-independent  intrachain distance as $\approx 2.5\,\mu$m which is by a factor of 20 larger than  $\lambda_L(0)\sim 0.1\,\mu$m. Since, according to the authors, the actual temperature at which the vortex structure forms might be  close to $T_c$, the relevant $\lambda_L$ is larger, so one may consider the above  estimate as having the correct order. It would be interesting to do the same experiment in a deliberately tilted field and compare the data with calculation based on particular elastic moduli of MgB$_2$.\\

 The author is grateful to J. Clem, S. Bud'ko,  R.~Prozorov,   M. Tanatar   for helpful   discussions. The Ames Laboratory is supported by the Department of Energy, Office of  Basic Energy Sciences, Division of Materials Sciences and Engineering under Contract No. DE-AC02-07CH11358.

\appendix

\section{Elasic moduli}

The tensor components are easily reproduced since  tensors transform as products of coordinates. In non-zero $\lambda$'s, index $x$ can appear even number of times, $z$ can come either in even numbers or in combination with $y$.

\begin{eqnarray}
\lambda_{xxxx} &=& \lambda_1  ,\nonumber\\
\lambda_{yyyy} &=&\lambda_1\cos^4\theta +\lambda_4\sin^4\theta +(\lambda_6+\lambda_5/2)  \sin^22\theta  , \nonumber\\   
\lambda_{xxyy} &=&\lambda_2\cos^2\theta + \lambda_5\sin^2\theta  \,,\nonumber\\
\lambda_{xyxy}&=&\lambda_3\cos^2\theta + \lambda_6\sin^2\theta  \,, \nonumber\\
\lambda_{xxzz} &=&\lambda_2\sin^2\theta + \lambda_5\cos^2\theta  \,,\nonumber\\
\lambda_{zzyy} &=& \lambda_5(\sin^4 \theta +\cos^4\theta)+(\lambda_1+\lambda_4-4\lambda_6)\sin^2\theta\cos^2\theta  \nonumber\\
\lambda_{xzxy} &=&(\lambda_3- \lambda_6)\sin \theta  \cos \theta  \,,\nonumber\\
\lambda_{xxyz} &=&(\lambda_2- \lambda_5)\sin \theta  \cos \theta  \,,\nonumber\\
\lambda_{yzyy} &=& [(\lambda_1- \lambda_5)\cos^2 \theta\nonumber\\
&+&(\lambda_5- \lambda_4)\sin^2 \theta -2 \lambda_6\cos 2 \theta\,]\sin \theta  \cos \theta  \,,\nonumber\\
\lambda_{yzzz} &=&( \lambda_1 \sin^2 \theta+3\lambda_5 \cos 2 \theta-\lambda_4\cos^2\theta)\sin \theta  \cos \theta  \,.\nonumber
\end{eqnarray}
Note that the last four entries are zeros for $\theta=0,\pi/2$. 

\section{Hexagonal cristals}

The elastic energy of hexagonal crystal  in the crystal frame is\cite{LL}
\begin{eqnarray}
&&F_h = (2 \lambda_{\xi\eta\xi\eta}+\lambda_{\xi\xi\eta\eta})(u_{xx}^2+u_{yy}^2 ) \nonumber\\ 
&&+ 2(2 \lambda_{\xi\eta\xi\eta}-\lambda_{\xi\xi\eta\eta})u_{xx} u_{yy} 
+4\lambda_{\xi\xi\eta\eta}u_{xy}^2 \label{Fhex}\\
&&+  \lambda_{zzzz}u_{zz}^2/2
+2 \lambda_{\xi\eta zz} (u_{xx}+u_{yy})u_{zz}
+4 \lambda_{\xi z\eta z} (u_{xz}^2+u_{yz}^2) \nonumber
\end{eqnarray}
 Here $ \lambda_{\xi\eta\xi\eta}, \lambda_{\xi\xi\eta\eta}, \lambda_{\xi\eta zz}, \lambda_{\xi z\eta z},  \lambda_{zzzz} $ are five independent elastic constants, $\xi=\eta^*=x+i y$. 
  
 Compare this with the energy $F_t$ for the tetragonal case which in the crystal frame and our notation reads:  
\begin{eqnarray}
&&F_t =   \lambda_1(u_{xx}^2+u_{yy}^2 )/2  
 +  \lambda_2u_{xx} u_{yy} 
+2\lambda_3u_{xy}^2
+  \lambda_4u_{zz}^2/2\nonumber\\
&&+2 \lambda_5 (u_{xx}+u_{yy})u_{zz}
+2 \lambda_6 (u_{xz}^2+u_{yz}^2)\,.
 \label{Ft} 
\end{eqnarray}
Clearly, $F_h=F_t$ if 
\begin{eqnarray}
&&  \lambda_1= 2(2 \lambda_{\xi\eta\xi\eta}+\lambda_{\xi\xi\eta\eta}) ,\,\,  
  \lambda_2=2(2 \lambda_{\xi\eta\xi\eta}-\lambda_{\xi\xi\eta\eta}) , \nonumber\\
&&  \lambda_3=2 \lambda_{\xi\xi\eta\eta}, \,\, 
 \lambda_4=\lambda_{zzzz},\,\, 
 \lambda_5 =\lambda_{\xi\eta zz}, \,\, 
  \lambda_6 =\lambda_{\xi  z\eta z}\,.\qquad
 \label{Ft=Fh} 
\end{eqnarray}
Hence, we have $\lambda_1=\lambda_2+2\lambda_3$.

\end{document}